\documentclass[sigconf,screen,table]{acmart}

\setcopyright{acmlicensed}
\acmDOI{10.1145/3650212.3652115}
\acmYear{2024}
\copyrightyear{2024}
\acmSubmissionID{issta24main-p316-p}
\acmISBN{979-8-4007-0612-7/24/09}
\acmConference[ISSTA '24]{Proceedings of the 33rd ACM SIGSOFT International Symposium on Software Testing and Analysis}{September 16--20, 2024}{Vienna, Austria}
\acmBooktitle{Proceedings of the 33rd ACM SIGSOFT International Symposium on Software Testing and Analysis (ISSTA '24), September 16--20, 2024, Vienna, Austria}
\received{15-DEC-2023}
\received[accepted]{2024-03-02}

\AtBeginDocument{%
  \providecommand\BibTeX{{%
    \normalfont B\kern-0.5em{\scshape i\kern-0.25em b}\kern-0.8em\TeX}}}

\usepackage{microtype}
\usepackage{booktabs}
\usepackage{caption}
\usepackage{balance}
\usepackage{subcaption}
\usepackage{graphicx}
\usepackage{graphbox}
\usepackage{hyperref}
\usepackage{setspace}
\usepackage{wrapfig}
\usepackage{makecell}
\usepackage{verbatim}
\usepackage{graphicx}
\usepackage{color}
\usepackage{url}
\usepackage{hyperref}
\usepackage{listings}
\usepackage{csquotes}
\usepackage[normalem]{ulem}
\usepackage{setspace}
\usepackage[ruled,linesnumbered]{algorithm2e}
\usepackage{threeparttable}
\usepackage[noend]{algpseudocode}
\usepackage{tablefootnote}
\usepackage{longtable}
\usepackage{amsthm}
\usepackage{fontawesome}
\usepackage{bbding}

\algblock{Input}{EndInput}
\algblock{Output}{EndOutput}

\usepackage{xparse}
\usepackage{mathtools}
\usepackage{multirow}
\usepackage{datapie}
\usepackage{adjustbox}
\usepackage[most]{tcolorbox}
\def\BibTeX{{\rm B\kern-.05em{\sc i\kern-.025em b}\kern-.08emT\kern-.1667em\lower.7ex\hbox{E}\kern-.125emX}}
\usepackage{soul}
\usepackage{caption}
\usepackage{enumitem}
\newcommand{\distance}{12pt}
\usepackage{wrapfig}

\newcommand{\done}[1]{}
\newcommand{\remove}[1]{}
\newcommand{\parabf}[1]{\noindent\textbf{#1}}
\newcommand{\mytextbf}[1]{\noindent\textbf{#1}}
\newcommand{\codeIn}[1]{\texttt{#1}}
\newcommand{\mybox}[1]{\begin{tcolorbox}[enhanced, frame hidden, boxsep=0pt]\emph{#1}\end{tcolorbox}}

\begin{document}
\title{CoderUJB: An Executable and Unified Java Benchmark for  Practical Programming Scenarios}

\author{Zhengran Zeng}
\orcid{0009-0009-8422-4522}
\affiliation{%
  \institution{Peking University}
  \city{Beijing}
  \country{China}
}
\email{zhengranzeng@stu.pku.edu.cn}

\author{Yidong Wang}
\orcid{0009-0007-9969-8259}
\affiliation{%
  \institution{Peking University}
  \city{Beijing}
  \country{China}
}
\email{yidongwang37@gmail.com}

\author{Rui Xie}
\orcid{0000-0002-1756-7746}
\affiliation{%
  \institution{Peking University}
  \city{Beijing}
  \country{China}
}
\authornote{Rui Xie, Wei Ye, and Shikun Zhang are the corresponding authors.}
\email{ruixie@pku.edu.cn}

\author{Wei Ye}
\orcid{0000-0002-9331-4716}
\affiliation{%
  \institution{Peking University}
  \city{Beijing}
  \country{China}
}
\authornotemark[1]
\email{wye@pku.edu.cn}

\author{Shikun Zhang}
\orcid{0000-0002-8576-2674}
\affiliation{%
  \institution{Peking University}
  \city{Beijing}
  \country{China}
}
\authornotemark[1]
\email{zhangsk@pku.edu.cn}

\begin{abstract}
In the evolving landscape of large language models (LLMs) tailored for software engineering, the need for benchmarks that accurately reflect real-world development scenarios is paramount. Current benchmarks are either too simplistic or fail to capture the multi-tasking nature of software development. To address this, we introduce CoderUJB, a new benchmark designed to evaluate LLMs across diverse Java programming tasks that are executable and reflective of actual development scenarios, acknowledging Java's prevalence in real-world software production. CoderUJB comprises 2,239 programming questions derived from 17 real open-source Java projects and spans five practical programming tasks. Our empirical study on this benchmark investigates the coding abilities of various open-source and closed-source LLMs, examining the effects of continued pre-training in specific programming languages code and instruction fine-tuning on their performance. The findings indicate that while LLMs exhibit strong potential, challenges remain, particularly in non-functional code generation (e.g., test generation and defect detection). Importantly, our results advise caution in the specific programming languages continued pre-training and instruction fine-tuning, as these techniques could hinder model performance on certain tasks, suggesting the need for more nuanced strategies. CoderUJB thus marks a significant step towards more realistic evaluations of programming capabilities in LLMs, and our study provides valuable insights for the future development of these models in software engineering.

\end{abstract}


\maketitle

\section{Introduction}

\begin{table}[]
\caption{Comparison of existing code benchmark and CoderUJB.}
\begin{adjustbox}{width=1\columnwidth}
\begin{tabular}{|c|c|c|c|c|}
\hline
\textbf{Benchmark} & \textbf{Questions} & \textbf{Executable} & \textbf{Project-Runnable} & \textbf{Multi-Tasks} \\ \hline\hline
HumanEval~\cite{codex}        & 164                  & \checkmark               & $\times$                        & $\times$                   \\ 
MultiPL-E~\cite{multiple}        & 2,952                & \checkmark                 & $\times$                        & $\times$                   \\ 
MBPP~\cite{mbpp}             & 974                  & \checkmark                 & $\times$                        & $\times$                   \\ 
CoderEval~\cite{codereval}        & 460                  & \checkmark                 & \checkmark                       & $\times$                   \\ 
Defects4j~\cite{defects4j}    & 835                  & \checkmark                 & \checkmark                       & $\times$                   \\ 
ChatTester~\cite{chatteser}       & 1,000                & \checkmark                 & \checkmark                       & $\times$                   \\ 
Libro~\cite{issuetest}            & 750                  & \checkmark                 & \checkmark                       & $\times$                   \\ 
CodeXGLUE~\cite{codexglue}        & 759,000              & $\times$                  & $\times$                        & \checkmark                  \\ 
XCodeEval~\cite{xcodeeval}        & 159,464              & \checkmark                 & $\times$                        & \checkmark                  \\ \hline\hline
CoderUJB         & 2,239                & \checkmark                 & \checkmark                       & \checkmark                  \\ \hline
\end{tabular}
\end{adjustbox}
\label{tab:benchmark}
\end{table}

Researchers have found that advanced AI technologies, exemplified by large language models (LLMs), are proficient in addressing a broad spectrum of challenges, spanning from everyday tasks to complex software engineering issues~\cite{chatgpt,LLMSurvey,codex,aprplm,codeagent1,wang2022usb,wang2023pandalm,wang2024exploring,kieval}. In addition to these general-purpose LLMs, there are code-centric large language models (code LLMs), such as CodeX~\cite{codex}, CodeLlama~\cite{codellama}, and StarCoder~\cite{starcoder}, which are specifically designed to excel at software engineering tasks. Many of these code LLMs are open-source and can be privately deployed to avoid data security issues. As a result, they have gained significant attention for their strong coding skills. Because of this interest, different benchmarks~\cite{codex,codereval,codexglue,xcodeeval,classeval,aixbench} have been designed to measure the programming capabilities of these LLMs. Specifically, Table~\ref{tab:benchmark} presents a selection of notable benchmarks within software engineering alongside our CoderUJB, illustrating their distinct characteristics. The HumanEval~\cite{codex} benchmark stands out in this field, which is used to evaluate the ability of Python function generation and consists of 164 manually designed Python programming questions, and evaluates the quality of the generated solutions by checking whether the solution can be successfully executed by unit tests. We denote those execution-based evaluations as "\textbf{Executable (\checkmark)}". Then, CoderEval~\cite{codereval} noticed that the questions in HumanEval are simple single-function generation tasks that do not match the actual development scenarios (i.e., writing code in a software project). So, they introduced a new benchmark with 460 questions that better aligned with the actual development scenarios. We denote those actual development questions as "\textbf{Project-Runnable (\checkmark)}", indicating that the generated code requires a project context dependency for execution. However, those benchmarks focus on a single programming scenario and cannot provide a comprehensive evaluation of the programming capability of LLMs. Previous multi-task benchmarks, like CodeXGLUE~\cite{codexglue}, have been critiqued~\cite{codex} because they rely on similarity-based metrics like BLEU and CodeBLEU~\cite{codebleu}, which do not involve running the code. The recently proposed multi-programming task dataset XCodeEval~\cite{xcodeeval} focuses on questions from programming competitions, which do not accurately reflect typical real-world development scenarios. Consequently, the absence of a benchmark that comprehensively covers multi-programming tasks, executability, and matches real-world development scenarios prevents us from assessing the effectiveness of current LLMs on a broader range of real-world programming tasks.

To this end, we introduce CoderUJB, a comprehensive benchmark designed for evaluating LLMs that supports multiple tasks, adheres to real-world software development scenarios, and allows for execution within a complete program context(i.e., all source code and execution environments). Specifically, CoderUJB is built on 17 real open-source Java projects, acknowledging Java's prevalence in real-world software production~\cite{tiobe}. We extracted 238 functional code generation questions and 140 code-based test generation~\cite{chatteser} questions from these projects by analyzing the abstract syntax trees and test coverage relationships of the project source code. Then, we extracted and collected 451 issue-based test generation~\cite{issuetest} questions, 470 automatic program repair~\cite{aprplm} questions, and 940 defect detection~\cite{devign} questions from the projects by combining the detailed defect information and related issue reports from the projects. Altogether, CoderUJB comprises 2,239 programming questions covering five trending and practical programming tasks, which is the largest benchmark that is "Project-Runnable" and each question comes with a complete program context to facilitate the researcher's detailed analysis of the questions and the generated solutions. Ultimately, the solutions generated by the LLMs will be placed in real projects for execution, and the programming ability of the LLMs will be evaluated using the execution success rate as the primary metric.

Next, to illustrate the value of CoderUJB to the field, we conducted a comprehensive empirical study on CoderUJB to explore the programming abilities of a representative set of open-source code LLMs and general-purpose closed-source LLMs. The aim was to answer a couple of crucial questions: How good are these LLMs at coding? And how do continued pre-training~\cite{codellama} and instruction fine-tuning~\cite{wizardcoder} affect their programming performance? After running a slew of experiments on CoderUJB, we find that current LLMs still perform poorly in solving non-functional code generation tasks, especially defect detection tasks. Secondly, superior open-source LLMs can already approach or even surpass GPT-3.5-Turbo on the functional code generation task and the two test generation tasks. Thirdly, continued pre-training in a specific programming language can reduce an LLM's performance in other languages, and this negative impact diminishes as the task becomes less related to the original pre-training task. Such varied outcomes across different tasks emphasize CoderUJB's value as a unified evaluation benchmark that incorporates multiple programming tasks. Lastly, our findings indicate that instruction fine-tuning diminishes the performance of code LLMs in functional code generation and the two test generation tasks—a contrast to the results from the less practical benchmark, HumanEval. This highlights the value of CoderUJB as a practical programming evaluation benchmark. 

We summarize the main contributions of this study as follows:

\begin{itemize}
    \item \textbf{Benchmark.} We have introduced CoderUJB, a universal Java benchmark for assessing LLM performance across multiple real-world programming tasks. The benchmark includes executable test cases and the complete program context (i.e., all source code and execution environments), comprising 2,239 programming questions.
    \item \textbf{Study.} We ran a comprehensive empirical analysis of both open-source and proprietary LLMs using CoderUJB, studying (1) their performance across diverse coding tasks, (2) the effect of continued pre-training in a specific language code, and (3) the impact of instruction fine-tuning on these models.
    \item \textbf{Implications.} This work revealed multiple significant findings: (1) Program context is useful in code generation tasks. (2) Caution is advised when continuing pre-training in a specific programming language, as its effects tend to be unpredictable, especially if the downstream task is substantially different from the pre-training task. (3) Instruction fine-tuning should be approached carefully, as it can diminish the performance of LLMs on tasks that align closely with the pre-trained task. (4) Comprehensive evaluations are essential, given that different programming tasks may yield disparate results when the same training strategy is applied. 
\end{itemize}

The CoderUJB are publicly available in \textbf{\url{https://github.com/WisdomShell/ujb}}.
\section{Background and Related Work}
\subsection{Large Language Models for Software Engineering}
\label{sec:2.1}

Recently, many code LLMs~\cite{codex,codegen,starcoder,codellama} have been proposed to solve software engineering tasks. These models typically leverage the Transformer Decoder~\cite{transformer} architecture and undergo extensive training on large-scale code databases~\cite{codex,codegen,starcoder}, expecting excellence in executing code-intensive tasks. 

A prominent example within this realm is CodeX~\cite{codex}. For the first time, they increased the model parameter size to 12 billion (B) and used 159 gigabytes (GB) of Python code data for pre-training. They also proposed the widely adopted benchmark HumanEval and the $pass@k$ metric for evaluating the quality of generated code, in which CodeX achieved a $pass@k\text{=}1$ of 28.81 on HumanEval. Subsequently, StarCoderBase~\cite{starcoder} further scaled both the parameter size to 15B and the training dataset to 799.37 GB code corpus, which contained various programming languages. We designate those LLMs that have not undergone specialized training for particular tasks (e.g., Python programming challenges or question-and-answer activities) as "\textbf{Base LLM}". 

As mentioned, to enhance Python-specific performance on tasks like HumanEval, researchers continued pre-training the base model StarCoderBase on an additional 35B tokens of Python code, yielding a Python-enhanced version of the model StarCoder-Python. Ultimately, StarCoderBase and StarCoder-Python achieved a $pass@k\text{=}1$ of 30.4 and 33.6 on HumanEval, respectively. We designate those LLMs that have continued pre-training in specific programming language corpus as "\textbf{Specific Programming Language (PL) Base}".

Moreover, the team behind WizardCoder~\cite{wizardcoder} found that fine-tuning base LLMs with high-quality instruction samples (i.e., more diverse and more challenging coding problems) further boosted the LLMs programming performance. They used an evolutionary instruction-based data collection strategy to collect 80k instruction fine-tuning samples from ChatGPT. Utilizing these novel datasets, they fine-tuned a "\textbf{Instruction Tuned}" LLM WizardCoder, significantly enhancing the HumanEval $pass@k\text{=}1$ rate from 33.57 to an impressive 58.12.  

Despite the progress, the prevailing methods of evaluating these code LLMs tend to focus on simple Python-based programming puzzles like HumanEval and lack a comprehensive assessment of other programming tasks in software engineering. This limited evaluation does not fully capture the advancements made in code LLMs, nor does it comprehensively evaluate the impact of specialized training processes such as continued pre-training in a specific language code (mostly in Python) and instructional fine-tuning on various practical software engineering scenarios.
In response to these limitations, our study introduces a multi-programming task, executable and real-world programming scenario-compliant evaluation benchmark, and provides an in-depth study to address the questions mentioned above.

\subsection{Coder Benchmark}
\label{sec:2.2}

Recent scholarly efforts have proposed many benchmarks~\cite{codex,mbpp,multiple,codereval,xcodeeval} to evaluate the programmatic skill of LLMs as presents in Table~\ref{tab:benchmark}.
The HumanEval~\cite{codex} benchmark stands out in this field, consisting of 164 manually designed Python programming questions. Each question provides function signatures, comments, function bodies, and multiple unit tests. LLMs are tasked with crafting function bodies informed by the given signatures and comments. Subsequently, the generated functions are executed and evaluated regarding their success in passing the unit tests tied to each question. We refer to such non-test-case code generation questions as functional code generation tasks. Later, CoderEval noticed that the questions
in HumanEval are simple "self-contained"~\cite{codereval} function generation tasks that each function only has language built-in dependency and do not match the actual development scenarios that rely on multiple public libraries and project files. Therefore, they introduced 460 Java and Python code generation questions derived from real projects on GitHub that are more aligned with actual development settings. 

In addition to mainstream functional code generation tasks, there are other types of datasets designed to evaluate LLMs. For example, Defects4j~\cite{defects4j}, a seminal benchmark in automated program repair, catalogs 835 defects across 17 authentic Java projects from Github. It offers both the defective and fixed versions of the code alongside related test cases. In this evaluative phase, the fix code generated by LLMs is executed, and the accompanying test cases verify the correctness of the fix. Based on Defects4j, Libro~\cite{issuetest} has developed a benchmark for generating issue-specific tests. This requires LLMs to generate test cases that trigger the corresponding defects based on a given issue report. Additionally, the ChatTester~\cite{chatteser} benchmark, designated for code-based test generation tasks, collects 1,000 test generation questions from open-source Java projects on GitHub. Each test question contains the code needed for testing, a natural language description of the task, and a test case. LLMs need to generate test cases based on the code under test and the natural language description. 

Besides the single-task benchmark mentioned above, previous researchers have proposed two multi-programming task benchmarks, CodeXGLUE~\cite{codexglue} and XCodeEval~\cite{xcodeeval}, that incorporate a broad range of questions and tasks, establishing a more substantial framework for evaluating LLMs. However, CodeXGLUE still uses textual similarity metrics, such as BLEU and CodeBLEU~\cite{codebleu}, thereby falling short of actual code execution. XCodeEval focuses on programming competition questions, which, like HumanEval, are not "Project-Runnable" and do not match the actual development scenarios. Therefore, CoderUJB fills the current gap in the field as a benchmark that simultaneously includes multiple programming tasks that are executable and match real-world development scenarios, thus broadening the scope of evaluation for LLMs across various practical programming tasks.

\section{CoderUJB Benchmark}
This section outlines the programming tasks included in the Coder -UJB and provides an overview of the dataset construction process and evaluation metrics.

\subsection{Tasks in CoderUJB}
CoderUJB has chosen four representative program generation tasks and one crucial program classification task to create a comprehensive and uniform dataset for evaluating programming skills. 

\parabf{Functional Code Generation (\textbf{FCG}):}
Code generation is a crucial topic for LLMs in software development~\cite{codex,codexglue}, often consuming much of the developer's time. In this task, we need to generate the corresponding function code based on the given function annotations. We refer to previous work~\cite{codereval,codex} to define the input of the task as function-related context, function annotations, and function signatures, while the output of the task is a function that meets the task requirements, and test cases are used to evaluate the quality of the generated function. Note that this task is limited to generating operational code that executes the service's logic (i.e., not test case function or configuration file).


\parabf{Code-based Test Generation (CTG):}
Test generation is another part of the code generation task, but it requires programming skills different from functional code generation. For example, writing test cases usually requires more cross-functional understanding, whereas generating functional functions tends to focus only on the content of the current function following decoupled design principles~\cite{llmtestsurvey}. Therefore, we refer to previous work~\cite{chatteser} on test generation as a separate task. This task involves reading and understanding the program logic of the code under test and then generating test cases that verify the code's core functionality based on the test cases' function annotations. Consequently, we define the input of the task as task-related context, test case annotations and test case signatures, and the output of the task as test cases that meet the task requirements, and evaluate the quality of the generated test cases based on metrics such as whether they meet the task requirements and test coverage, which will be detailed in Section~\ref{sec:3.5}.

\parabf{Issue-based Test Generation (\textbf{ITG}):}
Issue report-based test generation is a process where LLMs analyze and comprehend bugs reported in issue reports. Following the comprehension, they devise test cases to reproduce those issues. We follow the task definition developed in previous works~\cite{issuetest} for this kind of task. The task input consists of two elements: an issue report detailing a particular bug and the corresponding function signature linked to that bug, while the output of the task is defined as the test case that meets the task requirements.

\parabf{Defect Detection (\textbf{DD}):}
Defect detection is a critical activity in software engineering, significantly influencing the overall software development process~\cite{devign,deepfl}. The primary purpose of this task is to detect possible bugs in the software's code, including, but not limited to, syntax errors and potential runtime errors. 
Following the previous work's task definition~\cite{devign}, our task input for defect detection would include the specific function under scrutiny, and the output is a statement that indicates whether or not a bug exists.

\parabf{Automated Program Repair (\textbf{APR}):}
After successfully identifying issues within a program, the next step is to make corrections. In the task of automated program repair, LLMs are given samples of functions containing defects, and they are expected to fix these errors, returning functions that are free of defects. This task is crucial in software engineering, so we align our definition with previous task definitions~\cite{aprplm}. In this case, the task input includes task-related contexts and defective functions; the output should be fixed functions. Eventually, test cases are used to evaluate the correctness of the fixed functions.

\subsection{Datasets in CoderUJB}
\begin{figure*}
  \centering
  \includegraphics[width=0.95\linewidth]{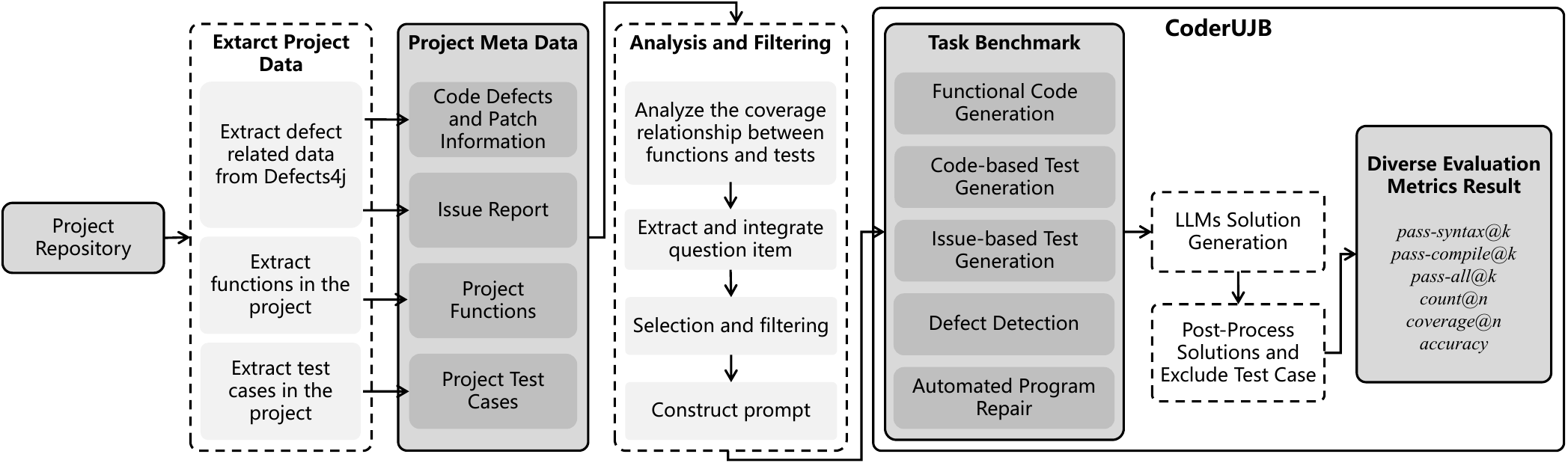}
  \caption{Overview of CoderUJB construction process.}
  \label{fig:coderujb}
\end{figure*}

This section introduces the process of collecting and processing CoderUJB questions. Figure~\ref{fig:coderujb} illustrates the overview of the data processing procedure within CoderUJB. As mentioned in Section~\ref{sec:2.2}, Defects4j~\cite{defects4j} provided 17 qualitative and practical open source Java projects as well as their defects data and issue reports, due to its completeness and quality, it has become the infrastructure in the field of automated program repair~\cite{aprplm,apr1,apr2}. Recognizing the value of Defects4j, we build a comprehensive benchmark on it. Consequently, each of the five programming tasks of CoderUJB gleaned its source data from 17 projects under Defects4j. Additionally, while it could incorporate other sources to build CodeUJB. we chose to collecting our coding problems across different tasks from the same reliable source (i.e., Defects4j), providing similar quality and difficulty levels across tasks and minimizing data quality impacts on our findings. Such selection offers substantial benefits over using varied datasets from disparate sources.

\subsection{Dataset Construction}
\parabf{Functional Code Generation (FCG):}
In the processing flow for the functional code generation benchmark, we first select the latest defect-free versions of the 17 open-source projects within Defects4j~\cite{defects4j} as the original project repository. We opt for the latest versions because they undergo continuous evolution and development, thus likely possessing the fewest potential defects and demonstrating superior code quality.
Subsequently, we extract all functions and test cases found in the project through an abstract syntax tree (AST) investigation. 
Then, as depicted in the "\codeIn{Analysis and Filtering Process}" in Figure~\ref{fig:coderujb}, we first analyze the coverage relationship between test cases and functions via a test coverage analysis tool~\cite{cobertura} to determine the test cases calling each function. Following this, we gather question-related data about each function, which includes the function body, comments, source code of the associated class, and related test cases. Later, we eliminated low-quality data by referring to CoderEval's guidelines~\cite{codereval}. More specifically, we ensure that:
1) The function is not a test, interface, or deprecated.
2) The function has function-level comment in English. 
3) The test coverage ratio larger than 50\%. 
Then, the question units are ranked in descending order based on the associated test coverage ratio, and a manual quality assessment is performed to further omit low-quality functions, such as oversimplified getter and setter functions. Ultimately, 238 functional code generation questions were filtered out, and each function had about 161.66 test cases on average. The question prompt is then formulated, and the exact structure will be discussed in detail in Section~\ref{sec:3.4}.


\parabf{Code-based Test Generation (CTG):}
For the code-based test generation benchmark, we also select the latest defect-free project version in Defects4j as the initial project repository. We then follow the same process for the functional code generation dataset to extract functions and test cases, while also examining their coverage relationship.
We subsequently filter out low-quality data using standards similar to those applied by CoderEval~\cite{codereval} and ChatTester~\cite{chatteser}. More specifically, we ensure that the test case 1) has name includes a test-associated keyword such as "Test", 2) is not deprecated, 3) has function-level comments in English, 4) is in a class that corresponds to a related functional class, such as "Example.java" and "ExampleTest.java".
Finally, we manually evaluate their quality to eliminate any low-quality instances, such as those with low test coverage, which are usually the bug reproduced test cases, and the ability to generate such test-cases we would like to examine in the following issue-based test generation task. We finally filtered out 140 code-based test cases to generate the test questions.



\parabf{Automated Program Repair (APR):}
As previously noted in Section~\ref{sec:2.2}, the Defects4j~\cite{defects4j} dataset is a prevalent source for automated program repair (APR). In designing our benchmark for APR, we directly adopt the defect dataset provided by Defects4j and refer to the previous work~\cite{aprplm} to focus only on the single-function defects in Defects4j. This specific focus makes it easier for us to create a uniform question prompt, which ultimately enhances the quality of our benchmark questions. In the end, we extracted 470 single-function defects from the dataset to serve as our APR benchmark. 

\parabf{Defect Detection (DD):}
For the defect detection benchmark, we have also developed a benchmark based on the Defects4j. Specifically, we focused on single-function defects as well, selecting 470 such bugs in Defects4j and using their defective functions along with their corresponding fixed versions as positive and negative samples. In the end, we have compiled a balanced dataset containing 940 detection samples for defect detection.

\parabf{Issue-based Test Generation (ITG):}
As mentioned in Section~\ref{sec:2.2}, LIBRO~\cite{issuetest} has proved a comprehensive benchmark for issue-based test generation based on Defects4j. Therefore, from the 750 test generation questions provided by LIBRO, we kept those related to issues that match the bugs in our automated program repair benchmark. Consequently, we obtained a benchmark consisting of 451 issue-based test generation questions. 

\subsection{Base Prompt Design}
\label{sec:3.4}

\begin{figure}
  \centering
  \begin{subfigure}[b]{1\columnwidth}
  \centering
  \includegraphics[width=\columnwidth]{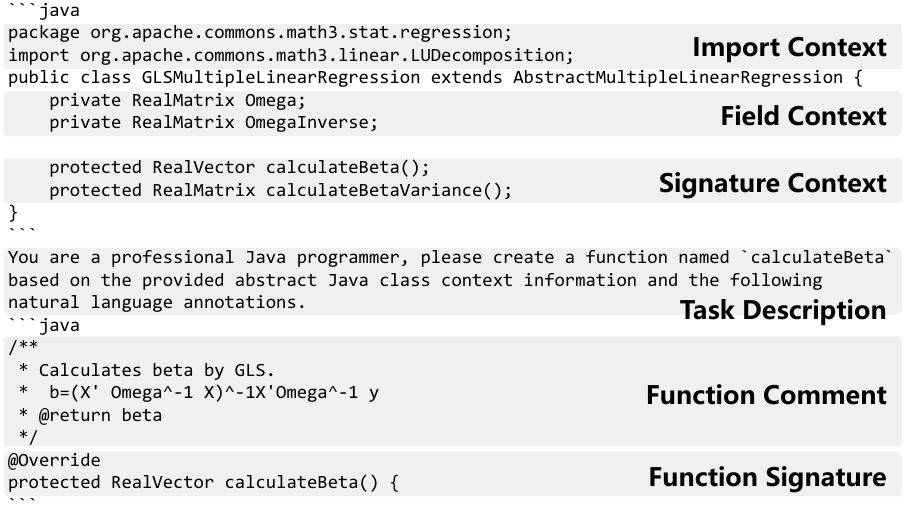}
  \caption{Prompt of chat invocation.}
  \label{fig:fcg_chat_prompt}
  \end{subfigure}
  \begin{subfigure}[b]{1\columnwidth}
  \centering
  \includegraphics[width=\columnwidth]{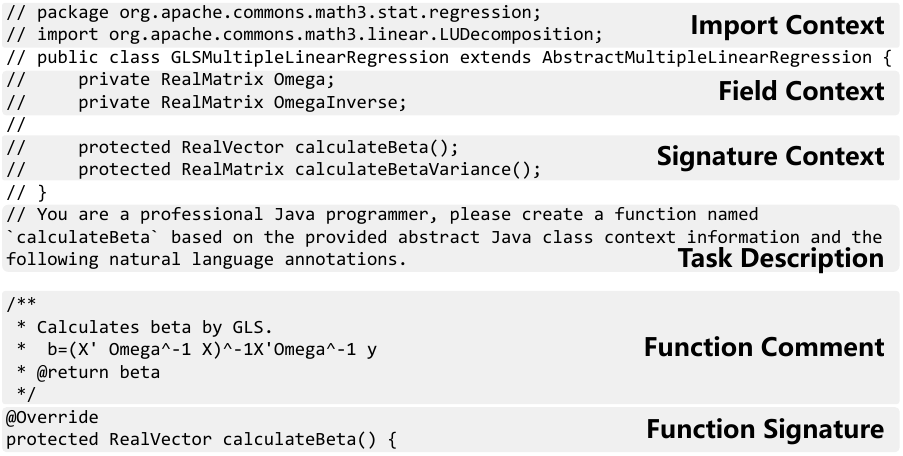}
  \caption{Prompt of complement invocation.}
  \label{fig:fcg_complete_prompt}
  \end{subfigure}
    \caption{Prompt of Functional Code Generation.}
    \label{fig:fcg_prompt}
\end{figure}

\parabf{Functional Code Generation (FCG):} Figure~\ref{fig:fcg_prompt} shows an example prompt for a functional code generation task consisting of six parts. Among them, "\codeIn{Import Context}", "\codeIn{Filed Context}", and "\codeIn{Signature Context}" serve as task-related contexts, extracted from the corresponding Java files of the task functions through AST analysis. We will show the effectiveness of those contexts in Section~\ref{sec:4.3.1}. The other three components, namely "\codeIn{Task Description}", "\codeIn{Function Comment}", and "\codeIn{Function Signature}", define the specific requirements of the task, instructing the LLMs on the desired code content to be generated.

It is important to note that the prevalent usage of current LLMs falls into two ways: chat invocation (e.g., ChatGPT~\cite{chatgpt} with chat alignment) and complement invocation (e.g., StarCoderBase~\cite{starcoder} without chat alignment). However, most users prefer the chat invocation method. Additionally, most open-source LLMs are typically offered only in the base version (i.e., without fine-tuning)~\cite{starcoder, codegen, llama2}, limiting us to using the model in a complement invocation way. For a fair comparison of these two types of LLMs, we have proposed two invocation prompts for each task: one in the style of a chat (i.e., Figure~\ref{fig:fcg_chat_prompt}) and the other in a complementary format (i.e., Figure~\ref{fig:fcg_complete_prompt}). While the content of the two prompts remains the same, they differ in how the information is formatted. Hence, we consider these two prompts equivalent, enabling a fair comparison between the two invocation ways.


\begin{figure}
  \centering
  \begin{subfigure}[b]{0.49\columnwidth}
  \centering
  \includegraphics[width=\textwidth]{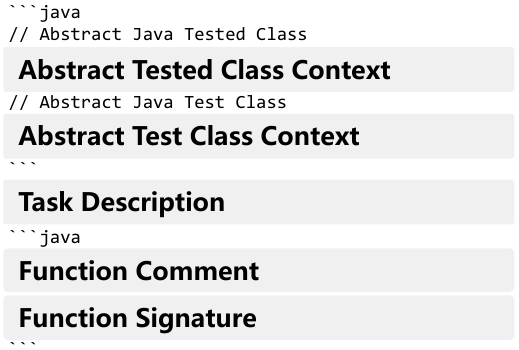}
  \caption{Chat prompt of Code-based Test Generation.}
  \label{fig:ctg_chat_prompt}
  \end{subfigure}
  \centering
  \begin{subfigure}[b]{0.49\columnwidth}
  \centering
  \includegraphics[width=\textwidth]{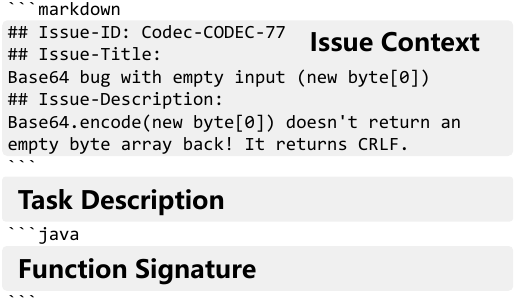}
  \caption{Chat prompt of Issue-based Test Generation.}
  \label{fig:itg_chat_prompt}
  \end{subfigure}
  \centering
  \begin{subfigure}[b]{0.49\columnwidth}
  \centering
  \includegraphics[width=\textwidth]{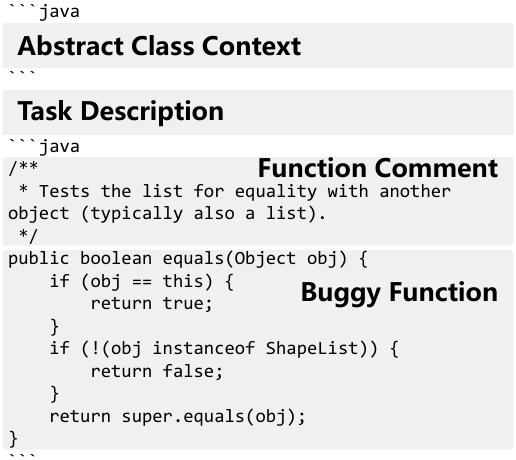}
  \caption{Chat prompt of Automated Program Repair.}
  \label{fig:apr_chat_prompt}
  \end{subfigure}
  \centering
  \begin{subfigure}[b]{0.49\columnwidth}
  \centering
  \includegraphics[width=\textwidth]{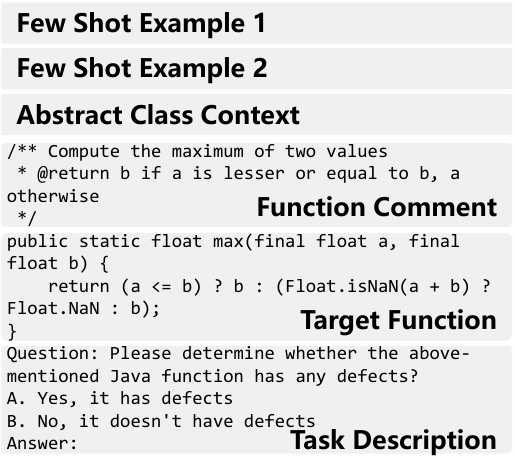}
  \caption{Complement prompt of Defect Detection.}
  \label{fig:dd_complete_prompt}
  \end{subfigure}
  
  \caption{Prompt of CoderUJB.}
  \label{fig:ucb_prompt}
\end{figure}

\parabf{Code-based Test Generation (CTG):} Figure~\ref{fig:ctg_chat_prompt} illustrates a chat invocation prompt for code-based test generation. In the task-related context, we provide both the under-test class context (i.e., the "\codeIn{Abstract Tested Class Context}" in Figure~\ref{fig:ctg_chat_prompt}) and the test class context (i.e., the "\codeIn{Abstract Test Class Context}" in Figure~\ref{fig:ctg_chat_prompt}). The term "\codeIn{Abstract Class Context}" here indicates the aggregate of "\codeIn{Import Context}", "\codeIn{Field Context}", and "\codeIn{Signature Context}" in Figure~\ref{fig:fcg_prompt}. Additionally, the task prompt also has a "\codeIn{Task Description}", "\codeIn{Function Comment}", and "\codeIn{Function Signature}" to outline the unique requirements of the task. Meanwhile, the complement invocation prompt for this task is similar to Figure~\ref{fig:fcg_chat_prompt}, with the prompt components reformatted and organized.

\parabf{Issue-based Test Generation (ITG):} Figure~\ref{fig:itg_chat_prompt} presents an example chat invocation prompt for issue-based test generation. This prompt format is guided by the LIBRO~\cite{issuetest}. More specifically, this task's prompt is composed of three key elements. The first element is the "\codeIn{Issue Context}"; it includes the issue report featuring the issue ID, title, and a detailed description specific to that issue. The other two elements, "\codeIn{Task Description}" and "\codeIn{Function Signature}" lay out the exact requirements of the task. The completion prompt for this task is also with the prompt components reformatted and organized. It is noteworthy that we do not provide "\codeIn{Abstract Class Context}" and "\codeIn{Function Comment}" in the prompt because we want to evaluate the capacity of LLMs in deriving those task information directly from issue reports~\cite{issuetest}.

\parabf{Automated Program Repair (APR):} Figure~\ref{fig:apr_chat_prompt} shows an example chat invocation prompt for an automated program repair task. The design of this prompt references from prior APR work~\cite{aprplm}. More specifically, this prompt has four key components and is similar to the design for functional code generation in Figure~\ref{fig:fcg_chat_prompt}. The main difference is that the prompt for the APR task includes the "\codeIn{Buggy Function}", which is the defective function that needs to be fixed by the LLMs.

\parabf{Defect Detection (DD):} 
Figure~\ref{fig:dd_complete_prompt} presents a complement invocation prompt for defect detection. Considering that this task is categorical, our prompt structure borrows from the design paradigms of classification benchmarks in the natural language processing (NLP) domain, such as MMLU~\cite{mmlu} and C-Eval~\cite{ceval}, which utilize few-shot examples to guide base LLMs (i.e., not fine-tuned with instructions) in generate valid output (i.e., the option of a multiple-choice question). We have implemented a two-shot format~\cite{fewshot}, following with relevant code context and a binary multiple-choice question.


\subsection{Metrics in CoderUJB}
\label{sec:3.5}
\subsubsection{$pass@k$}
We apply the $pass@k$ metric to evaluate the generated solutions of the code generation task as it has been widely used in previous researches~\cite{codex,codereval,nl2fix}. Specifically, given an unordered set of $n$ candidate solutions, $pass@k$ indicates the probability of selecting at least one correct solution in $k$ solutions sample from all $n$ candidate solutions. We use the following formula to compute $pass@k$ defined by previous work~\cite{codex}:
\begin{align}
pass@k := \underset{problems}{\mathbb{E}} \left [ 1-\frac{\binom{n-c}{k} }{\binom{n}{k} } \right ] 
\end{align}
Given that the number of generated solutions is $n$, the number of solutions used to estimate $pass@k$ is $k$, and $c$ is the number of correct solutions out of $n$ samples. Moreover, the $pass@k$ result for the entire dataset is the expected value (mean) of the $pass@k$ for individual problems. It is important to note that we have set $n$ to 20 in our study (doubling the size used in prior research~\cite{codereval}) to improve the stability of our findings.

In functional code generation and automated program repair scenarios, a solution is correct if it passes all relevant tests. In the test generation scenario, we refer to the definition of previous work~\cite{chatteser} and define a solution to be correct if and only if the test case can be run successfully and covers the tested code. In addition, in the issue-based test generation scenario, the test case must reproduce the issue bug by reporting the error in the defective project version but running successfully in the fixed version~\cite{issuetest}. 
Beyond executing correct metrics, we recorded the outcomes at various executing stages to better examine the solution. Specifically, we employ a $pass@k$ metric for syntax checking, defined as $pass\text{-}syntax@k$, where correctness equates to successful syntax checking, and another metric used for compilation checking, referred to as $pass\text{-}compile@k$, where we consider the solution to be correct if it compiles successfully. Lastly, we use a metric, $pass\text{-}all@k$, for situations where all test cases and checklists are passed.

\subsubsection{$count@n$}
To demonstrate the ability of LLMs to solve code-generation problems more intuitively, we introduce a metric $count$ -$@n$, which is similar to the standard metric $plausible patches$ in the field of APR~\cite{aprplm,nl2fix,apr1,apr2}. Specifically, this metric measures the number of coding problems an LLM can successfully solve by generating $n$ solutions for each problem. Therefore, the $count@n$ value assigned to each problem is defined as: 
\begin{align}
count@n := \bigvee_{i=1}^{n}correct(solution_i)
\end{align}
If any of the $n$ solutions generated by the LLMs for a given problem is correct, the $count@n$ value for that problem is assigned as 1, and for all other situations, it would be 0. The $count@n$ score across an entire benchmark is the sum of $count@n$ values for each problem within the benchmark.

\subsubsection{$coverage@n$}
Test coverage is a crucial metric for measuring the effectiveness of test cases~\cite{llmtestsurvey}. Therefore, we also counted the combined test coverage for the code under test for the test generation task. Specifically, for each set of $n$ solutions generated for a particular programming problem, we first identify and record the lines of code that each solution can cover. Subsequently, we accumulate the code lines covered from all $n$ solutions into a comprehensive set. The final step is to compute the percentage of this comprehensive set's lines of code against the total lines of code under test. The $coverage@n$ of the entire benchmark is then derived as the mean of the $coverage@n$ for each problem within the benchmark. Consequently, we define the $coverage@n$ metric as follow: 
\begin{align}
coverage@n := \frac{count(\bigcup\limits_{i = 1}^N{cover\_line(solution_i))}}{all\_tested\_line\_count}
\end{align}
The $cover\_line$ function can get the lines of code that a specific solution $solution_i$ can cover, and $all\_tested\_line$ indicates the number of lines of all tested code in the programming problem.

\subsubsection{$accuracy$}
For the classification task of defect detection, we simply adopt the widely used $accuracy$ metrics~\cite{mmlu,codexglue,ceval}.

\section{The extensive study}
In this section, we further evaluate existing leading LLMs with CoderUJB to delve into issues that are pertinent to researchers and to showcase CoderUJB's contribution to advancing the field.

\subsection{Research Questions}
This study investigates the following research questions:
\begin{itemize}[leftmargin=*]
    \item \parabf{RQ1:} \emph{Does the basic program context prompt improve the performance of LLMs on CoderUJB?} To this research question, we aim to explore how various prompts influence the performance of LLMs to confirm whether the basic program context given by CoderUJB is beneficial.
    
    \item \parabf{RQ2:} \emph{How do open-source and closed-source LLMs perform under CoderUJB?} We are looking to better understand the current progress of two types of LLMs and provide a thorough evaluation of their effectiveness in handling real programming tasks.

    \item \parabf{RQ3:} \emph{How does continued pre-training of specific programming language (PL) data affect the performance of code LLMs under CoderUJB?} This part of our study will explore how continued pre-training of specific programming language data might impact the performance of code LLMs when handling real programming tasks in other coding languages.

    \item \parabf{RQ4:} \emph{How does instruction fine-tuning influence the performance of code LLMs in CoderUJB?} This research question will investigate the potential benefits of instruction fine-tuning strategies on the performance of code LLMs under CoderUJB to provide feasible guidelines for the practical application of code LLMs.
\end{itemize}

\subsection{Code LLMs Subjects}
\begin{table}[]
\caption{Statistical information of the studied LLMs.}
\begin{adjustbox}{width=1.0\columnwidth}
\begin{tabular}{|c|c|c|c|c|}
\hline
\textbf{Type}                                                                & \textbf{Model Name} & \textbf{Size (B)} & \textbf{Trained From} & \textbf{Training Dataset} \\ \hline\hline
\multirow{2}{*}{Base}                                                        & CodeLlama           & 7;13;34           & Llama2                & 520B code tokens        \\ 
                                                                             & StarCoder-Base      & 15                & From Scratch          & 1T code tokens          \\
                                                                             & CodeShell      & 7                & From Scratch          & 500B code tokens          \\ \hline
\multirow{3}{*}{\begin{tabular}[c]{@{}c@{}}Specific PL\\ Base\end{tabular}}  & CodeLlama-Python    & 7;13;34           & CodeLlama            & 120B Python tokens      \\ 
                                                                             & StarCoder-Python    & 15                & StarCoder-Base        & 35B Python tokens       \\ 
                                                                             & StarCoder-Java      & 15                & StarCoder-Base        & 35B Java tokens         \\ \hline
\multirow{3}{*}{\begin{tabular}[c]{@{}c@{}}Instruction\\ Tuned\end{tabular}} & CodeLlama-Instruct  & 7;13;34           & CodeLlama             & 5B instruction tokens   \\ 
                                                                             & WizardCoder-Python  & 7;13;34           & CodeLlama-Python      & 80k instructions        \\ 
                                                                             & WizardCoder         & 15                & StarCoder-Python      & 80k instructions        \\
                                                                             & CodeShell-Chat         & 7                & CodeShell      & 40k instructions        \\ \hline
\multirow{3}{*}{\begin{tabular}[c]{@{}c@{}}Closed\\ Source\end{tabular}}     & Claude-1            & /                 & /                     & /                       \\ 
                                                                             & GPT-3.5-Turbo       & /                 & /                     & /                       \\ 
                                                                             & GPT-4               & /                 & /                     & /                       \\ \hline
\end{tabular}
\end{adjustbox}
\label{tab:llms}
\end{table}

For the study subject, we focus on the widely used code LLMs and three closed-source commercial LLMs. Table~\ref{tab:llms} provides an overview of these selected models. Specifically, take CodeLlama-7B as an example, "Trained From" and "Training Dataset" represent CodeLlama-7B is trained from Llama2-7B with 500B tokens code corpus and 20B tokens long context corpus. We have classified these LLMs into four primary categories: 

\mytextbf{Base LLMs:} This type of LLM consists of the widely adopted CodeLlama-7B, 13B, 34B~\cite{codellama}, and StarCoderBase-15B~\cite{starcoder}. We chose these models because they each come with their own Specific-PL-Base and Instruction-Tuned versions, aiding our future comparison studies and experiments. Meanwhile, previous studies~\cite{wizardcoder,pangu,codestudy1,codestudy2} have thoroughly researched these models, making them noteworthy representatives of code LLMs. Additionally, we add another CodeShell~\cite{codeshell} as a baseline for LLM with lower training resources, as it is trained from scratch using only 500B code tokens. We apply complement prompt for interacting with those LLMs.

\mytextbf{Specific-PL-Base LLMs:} In addition to the Base LLMs, many existing code LLMs~\cite{starcoder,codellama,codegen} have an additional version that undergoes further pre-training on Python data to improve the performance of the base LLM under Python programming tasks. However, this can raise concerns for researchers about how these models would perform with tasks in other programming languages~\cite{multiple,codegeex}. Therefore, we collected four specific Python base LLMs, namely CodeLlama-Python-7B, 13B, 34B~\cite{codellama}, and StarCoder-15B~\cite{starcoder}. Moreover, to investigate how continued pre-training on specific programming languages data would affect LLMs on programming tasks in other languages, we follow the setting of StarCoder-Python to further continue pre-training StarCoder-Base on another 35B Java tokens (random sampling from The Stack~\cite{thestack}) and get a specific Java model StarCoder-Java. We apply complement prompt for interacting with those LLMs. 

\mytextbf{Instruction-Tuned LLMs:} Along with the Base LLMs, instruction fine-tuned LLMs are another crucial category of interest~\cite{instructgpt,wizardcoder}. Unlike traditional fine-tuning, which focuses on single-task training, instruction fine-tuning employs diverse task data to train the model. Previous studies~\cite{instructgpt,wizardcoder,codestudy1,pangu} have shown that instruction fine-tuning enhances performance across various NLP and programming tasks, showing more promise than traditional fine-tuning strategy. With this in mind, we selected eight open-source, instruction fine-tuned LLMs named CodeLlama-Instruct-7B, 13B, 34B~\cite{codellama}, WizardCoder-Python-7B, 13B, 34B, WizardCoder-15B~\cite{wizardcoder}, and CodeShell-Chat~\cite{codeshell} for our exploration of how instruction fine-tuning influences LLMs in distinct programming tasks. We apply chat prompt for interacting with those LLMs.

\mytextbf{Closed-Source LLMs:} We also select three closed-source commercial LLMs (i.e., Claude-1~\cite{claude}, GPT-3.5-Turbo-0301~\cite{chatgpt}, and GPT-4-0314~\cite{gpt4}) to evaluate the gap between open-source and closed-source LLMs. We apply chat prompt for interacting with those LLMs.


\subsection{Results and Analysis}

\subsubsection{RQ1: Does the Basic Program Context Prompt Improve the Performance of LLMs on CoderUJB}
\label{sec:4.3.1}
\begin{table}[]
\caption{Evaluation results of prompt design for CoderUJB.}
\begin{adjustbox}{width=1\columnwidth}
\begin{tabular}{|c|c|ccc|ccc|}
\hline
\multirow{2}{*}{\textbf{Task}} & \multirow{2}{*}{\textbf{Metric}} & \multicolumn{3}{c|}{\textbf{StarCoderBase-15B}}                                                                                                                                                                                           & \multicolumn{3}{c|}{\textbf{CodeLlama-13B}}                                                                                                                                                                                               \\ \cline{3-8} 
                               &                                  & \multicolumn{1}{c}{\textbf{\begin{tabular}[c]{@{}c@{}}$program$\\ $context$\end{tabular}}} & \multicolumn{1}{c}{\textbf{\begin{tabular}[c]{@{}c@{}}$one$\\ $shot$\end{tabular}}} & \textbf{\begin{tabular}[c]{@{}c@{}}$four$\\ $shot$\end{tabular}} & \multicolumn{1}{c}{\textbf{\begin{tabular}[c]{@{}c@{}}$program$\\ $context$\end{tabular}}} & \multicolumn{1}{c}{\textbf{\begin{tabular}[c]{@{}c@{}}$one$\\ $shot$\end{tabular}}} & \textbf{\begin{tabular}[c]{@{}c@{}}$four$\\ $shot$\end{tabular}} \\ \hline\hline
\multirow{3}{*}{\textbf{FCG}}  & $pass\text{-}all@k\text{=}1$     & \multicolumn{1}{c}{\textbf{15.32}}                                                     & \multicolumn{1}{c}{12.18}                                                       & 11.66                                                        & \multicolumn{1}{c}{\textbf{21.91}}                                                     & \multicolumn{1}{c}{12.84}                                                       & 14.50                                                        \\ 
                               & $pass\text{-}all@k\text{=}10$    & \multicolumn{1}{c}{\textbf{26.82}}                                                     & \multicolumn{1}{c}{19.33}                                                       & 19.66                                                        & \multicolumn{1}{c}{\textbf{34.85}}                                                     & \multicolumn{1}{c}{20.92}                                                       & 22.37                                                        \\  
                               & $count\text{-}all@n\text{=}20$   & \multicolumn{1}{c}{\textbf{75}}                                                        & \multicolumn{1}{c}{51}                                                          & 53                                                           & \multicolumn{1}{c}{\textbf{90}}                                                        & \multicolumn{1}{c}{55}                                                          & 57                                                           \\ \hline
\multirow{4}{*}{\textbf{CTG}}  & $pass\text{-}all@k\text{=}1$     & \multicolumn{1}{c}{\textbf{12.14}}                                                     & \multicolumn{1}{c}{3.93}                                                        & 5.57                                                         & \multicolumn{1}{c}{\textbf{12.61}}                                                     & \multicolumn{1}{c}{7.04}                                                        & 7.57                                                         \\  
                               & $pass\text{-}all@k\text{=}10$    & \multicolumn{1}{c}{\textbf{31.80}}                                                     & \multicolumn{1}{c}{12.24}                                                       & 11.26                                                        & \multicolumn{1}{c}{\textbf{38.17}}                                                     & \multicolumn{1}{c}{17.50}                                                       & 17.69                                                        \\  
                               & $count\text{-}all@n\text{=}20$   & \multicolumn{1}{c}{\textbf{52}}                                                        & \multicolumn{1}{c}{24}                                                          & 20                                                           & \multicolumn{1}{c}{\textbf{67}}                                                        & \multicolumn{1}{c}{30}                                                          & 29                                                           \\  
                               & $coverage@n\text{=}20$           & \multicolumn{1}{c}{\textbf{11.91}}                                                     & \multicolumn{1}{c}{6.20}                                                        & 5.39                                                         & \multicolumn{1}{c}{\textbf{15.66}}                                                     & \multicolumn{1}{c}{8.65}                                                        & 8.52                                                         \\ \hline
\multirow{3}{*}{\textbf{APR}}  & $pass\text{-}all@k\text{=}1$     & \multicolumn{1}{c}{\textbf{6.56}}                                                      & \multicolumn{1}{c}{4.14}                                                        & 4.64                                                         & \multicolumn{1}{c}{\textbf{4.50}}                                                      & \multicolumn{1}{c}{4.07}                                                        & 4.10                                                         \\  
                               & $pass\text{-}all@k\text{=}10$    & \multicolumn{1}{c}{\textbf{12.54}}                                                     & \multicolumn{1}{c}{7.69}                                                        & 8.11                                                         & \multicolumn{1}{c}{\textbf{8.39}}                                                      & \multicolumn{1}{c}{8.20}                                                        & 7.47                                                         \\  
                               & $count\text{-}all@n\text{=}20$   & \multicolumn{1}{c}{\textbf{66}}                                                        & \multicolumn{1}{c}{41}                                                          & 44                                                           & \multicolumn{1}{c}{44}                                                                 & \multicolumn{1}{c}{\textbf{45}}                                                 & 39                                                           \\ \hline
\multirow{2}{*}{\textbf{DD}}   & $accuracy$                       & \multicolumn{1}{c}{50.32}                                                              & \multicolumn{1}{c}{\textbf{51.19}}                                              & 48.70                                                        & \multicolumn{1}{c}{48.60}                                                               & \multicolumn{1}{c}{49.68}                                                       & \textbf{51.51}                                               \\  
                               & $error\text{-}count$             & \multicolumn{1}{c}{0}                                                                & \multicolumn{1}{c}{\textbf{0}}                                                  & 0                                                            & \multicolumn{1}{c}{32}                                                                & \multicolumn{1}{c}{\textbf{2}}                                                  & 4                                                            \\ \hline
\end{tabular}
\end{adjustbox}
\label{tab:eval-prompt}
\end{table}

\begin{table*}[]
\caption{Evaluation results for CoderUJB, HumanEval and CoderEval. $pass\text{-}all@k$ denoted as $p\text{-}a\text{=}k$, $count\text{-}all@n$ denoted as $c\text{-}a\text{=}n$, $accuracy$ denoted as $acc$, $error\text{-}count$ denoted as $err$. The values in {\color[HTML]{C00000} Red} indicate underperform, {\color[HTML]{00B050} Green} values indicate outperform corresponding "Trained From" LLMs as presents in Table~\ref{tab:llms}.}
\begin{adjustbox}{width=1\textwidth}
\begin{tabular}{|c|c|cccccccccccccc|cc|cc|}
\hline
                                                                                              &                                     & \multicolumn{14}{c|}{\textbf{CoderUJB}}                                                                                                                                                                                                                                                                                                                                                                                                                                                                  & \multicolumn{2}{c|}{\textbf{HumanEval}\tablefootnote{The results adopt from self-report value and Humaneval leader-board~\cite{bigcode-evaluation-harness} }}                     & \multicolumn{2}{c|}{\textbf{CoderEval}\tablefootnote{The results were obtained through our own execution of the official evaluation scripts.}}                     \\ \cline{3-20} 
                                                                                              &                                     & \multicolumn{3}{c|}{\textbf{FCG}}                                                                            & \multicolumn{3}{c|}{\textbf{CTG}}                                                                            & \multicolumn{3}{c|}{\textbf{ITG}}                                                                           & \multicolumn{3}{c|}{\textbf{APR}}                                                                            & \multicolumn{2}{c|}{\textbf{DD}}              & \textbf{Java}                & \textbf{Py}                  & \textbf{Java}                & \textbf{Py}                  \\ \cline{3-20} 
                                                                                              &                                     & \multicolumn{2}{c}{\textbf{$p\text{-}a@k$}}                 & \multicolumn{1}{c|}{\textbf{$c\text{-}a@n$}}   & \multicolumn{2}{c}{\textbf{$p\text{-}a@k$}}                 & \multicolumn{1}{c|}{\textbf{$c\text{-}a@n$}}   & \multicolumn{2}{c}{\textbf{$p\text{-}a@k$}}                & \multicolumn{1}{c|}{\textbf{$c\text{-}a@n$}}   & \multicolumn{2}{c}{\textbf{$p\text{-}a@k$}}                & \multicolumn{1}{c|}{\textbf{$c\text{-}a@n$}}    & \textbf{$acc$}        & \textbf{$err$}        & \multicolumn{2}{c|}{\textbf{$p\text{-}a@k$}}                & \multicolumn{2}{c|}{\textbf{$p\text{-}a@k$}}                \\ \cline{3-20} 
\multirow{-4}{*}{\textbf{Type}}                                                               & \multirow{-4}{*}{\textbf{Model-ID}} & \textbf{$k\text{=}1$}        & \textbf{$k\text{=}10$}       & \multicolumn{1}{c|}{\textbf{$n\text{=}20$}}    & \textbf{$k\text{=}1$}        & \textbf{$k\text{=}10$}       & \multicolumn{1}{c|}{\textbf{$n\text{=}20$}}    & \textbf{$k\text{=}1$}       & \textbf{$k\text{=}10$}       & \multicolumn{1}{c|}{\textbf{$n\text{=}20$}}    & \textbf{$k\text{=}1$}       & \textbf{$k\text{=}10$}       & \multicolumn{1}{c|}{\textbf{$n\text{=}20$}}     & \textbf{$n\text{=}1$} & \textbf{$n\text{=}1$} & \textbf{$k\text{=}1$}        & \textbf{$k\text{=}1$}        & \textbf{$k\text{=}1$}        & \textbf{$k\text{=}1$}        \\ \hline\hline
                                                                                              & \textbf{CodeShell-7B}               & 9.68                        & 16.83                        & \multicolumn{1}{c|}{45}                        & 6.82                        & 20.11                        & \multicolumn{1}{c|}{33}                        & 4.12                        & 7.29                        & \multicolumn{1}{c|}{49}                        & 3.59                        & 8.58                         & \multicolumn{1}{c|}{46}                         & 47.62                 & 43                    & 30.43                        & 34.30                        & 24.63                        & 19.78                        \\
                                                                                              & \textbf{CodeLlama-7B}               & 15.06                        & 25.00                        & \multicolumn{1}{c|}{65}                        & 10.79                        & 29.72                        & \multicolumn{1}{c|}{48}                        & 4.32                        & 10.96                        & \multicolumn{1}{c|}{61}                        & 3.66                        & 7.63                         & \multicolumn{1}{c|}{40}                         & 46.54                 & 25                    & 29.20                        & 29.98                        & 31.26                        & 24.08                        \\
                                                                                              & \textbf{CodeLlama-13B}              & 21.91                        & 34.85                        & \multicolumn{1}{c|}{90}                        & 12.61                        & 38.17                        & \multicolumn{1}{c|}{67}                        & 6.14                        & 13.83                        & \multicolumn{1}{c|}{71}                        & 4.50                        & 8.39                         & \multicolumn{1}{c|}{44}                         & 48.60                 & 32                    & 32.23                        & 35.07                        & 35.02                        & 23.73                        \\
                                                                                              & \textbf{CodeLlama-34B}              & 22.82                        & 36.54                        & \multicolumn{1}{c|}{96}                        & 14.57                        & 32.07                        & \multicolumn{1}{c|}{52}                        & \textbf{7.34}               & 14.16                        & \multicolumn{1}{c|}{73}                        & 5.01                        & 8.34                         & \multicolumn{1}{c|}{44}                         & 48.16                 & 27                    & 40.19                        & 45.11                        & 33.00                        & 27.23                        \\
\multirow{-4}{*}{\textbf{Base}}                                                               & \textbf{StarCoderBase-15B}          & 15.32                        & 26.82                        & \multicolumn{1}{c|}{75}                        & 12.14                        & 31.80                        & \multicolumn{1}{c|}{52}                        & 6.32                        & 12.58                        & \multicolumn{1}{c|}{64}                        & 6.56                        & 12.54                        & \multicolumn{1}{c|}{66}                         & 50.32                 & 0                     & 28.53                        & 30.35                        & 30.58                        & 21.34                        \\ \hline
                                                                                              & \textbf{CodeLlama-Python-7B}        & {\color[HTML]{C00000} 10.69} & {\color[HTML]{C00000} 21.70} & \multicolumn{1}{c|}{{\color[HTML]{C00000} 59}} & {\color[HTML]{00B050} 12.00} & {\color[HTML]{C00000} 29.06} & \multicolumn{1}{c|}{48}                        & {\color[HTML]{C00000} 4.08} & {\color[HTML]{C00000} 8.73}  & \multicolumn{1}{c|}{{\color[HTML]{C00000} 46}} & {\color[HTML]{C00000} 3.23} & {\color[HTML]{C00000} 6.96}  & \multicolumn{1}{c|}{40}                         & 49.46                 & 12                    & {\color[HTML]{C00000} 29.15} & {\color[HTML]{00B050} 40.48} & {\color[HTML]{C00000} 26.41} & {\color[HTML]{00B050} 24.56} \\
                                                                                              & \textbf{CodeLlama-Python-13B}       & {\color[HTML]{C00000} 14.43} & {\color[HTML]{C00000} 26.93} & \multicolumn{1}{c|}{{\color[HTML]{C00000} 72}} & {\color[HTML]{00B050} 13.00} & {\color[HTML]{C00000} 31.68} & \multicolumn{1}{c|}{{\color[HTML]{C00000} 54}} & {\color[HTML]{C00000} 3.08} & {\color[HTML]{C00000} 7.50}  & \multicolumn{1}{c|}{{\color[HTML]{C00000} 41}} & {\color[HTML]{C00000} 3.72} & {\color[HTML]{C00000} 7.07}  & \multicolumn{1}{c|}{{\color[HTML]{C00000} 39}}  & 47.41                 & 37                    & {\color[HTML]{00B050} 33.56} & {\color[HTML]{00B050} 42.89} & {\color[HTML]{C00000} 28.11} & {\color[HTML]{00B050} 26.23} \\
                                                                                              & \textbf{CodeLlama-Python-34B}       & {\color[HTML]{C00000} 14.31} & {\color[HTML]{C00000} 26.23} & \multicolumn{1}{c|}{{\color[HTML]{C00000} 72}} & {\color[HTML]{C00000} 11.71} & {\color[HTML]{C00000} 28.24} & \multicolumn{1}{c|}{{\color[HTML]{C00000} 48}} & {\color[HTML]{C00000} 5.32} & {\color[HTML]{C00000} 11.61} & \multicolumn{1}{c|}{{\color[HTML]{C00000} 63}} & {\color[HTML]{00B050} 5.54} & {\color[HTML]{00B050} 9.27}  & \multicolumn{1}{c|}{{\color[HTML]{00B050} 46}}  & 49.24                 & 15                    & {\color[HTML]{C00000} 39.46} & {\color[HTML]{00B050} 53.29} & {\color[HTML]{C00000} 29.17} & {\color[HTML]{C00000} 24.73} \\
                                                                                              & \textbf{StarCoder-Python-15B}       & {\color[HTML]{C00000} 14.39} & {\color[HTML]{C00000} 25.97} & \multicolumn{1}{c|}{{\color[HTML]{C00000} 69}} & {\color[HTML]{C00000} 7.75}  & {\color[HTML]{C00000} 26.02} & \multicolumn{1}{c|}{{\color[HTML]{C00000} 44}} & {\color[HTML]{00B050} 6.82} & {\color[HTML]{00B050} 13.34} & \multicolumn{1}{c|}{{\color[HTML]{00B050} 68}} & {\color[HTML]{00B050} 8.37} & {\color[HTML]{00B050} 14.42} & \multicolumn{1}{c|}{{\color[HTML]{00B050} 74}}  & 50.54                 & 1                     & {\color[HTML]{00B050} 30.22} & {\color[HTML]{00B050} 33.57} & {\color[HTML]{C00000} 29.26} & {\color[HTML]{00B050} 21.46} \\
\multirow{-5}{*}{\textbf{\begin{tabular}[c]{@{}c@{}}Specific\\ Language\\ Base\end{tabular}}} & \textbf{StarCoder-Java-15B}         & {\color[HTML]{00B050} 18.82} & {\color[HTML]{00B050} 30.28} & \multicolumn{1}{c|}{{\color[HTML]{00B050} 77}} & {\color[HTML]{C00000} 10.43} & {\color[HTML]{00B050} 32.05} & \multicolumn{1}{c|}{{\color[HTML]{00B050} 56}} & {\color[HTML]{C00000} 5.75} & {\color[HTML]{C00000} 11.11} & \multicolumn{1}{c|}{{\color[HTML]{C00000} 59}} & {\color[HTML]{C00000} 6.27} & {\color[HTML]{C00000} 10.82} & \multicolumn{1}{c|}{{\color[HTML]{C00000} 57}}  & 49.14                 & 3                     & {\color[HTML]{00B050} 30.62} & {\color[HTML]{C00000} 27.07} & {\color[HTML]{00B050} 31.84} & {\color[HTML]{C00000} 14.89} \\ \hline
                                                                                              & \textbf{CodeShell-Chat-7B}      & {\color[HTML]{C00000} 7.79} & {\color[HTML]{C00000} 15.66} & \multicolumn{1}{c|}{{\color[HTML]{C00000} 43}}                        & {\color[HTML]{C00000} 2.82}  & {\color[HTML]{C00000} 10.97} & \multicolumn{1}{c|}{{\color[HTML]{C00000} 18}} & {\color[HTML]{C00000} 3.26} & {\color[HTML]{00B050} 7.42}  & \multicolumn{1}{c|}{{\color[HTML]{C00000} 38}} & {\color[HTML]{00B050} 3.79} & {\color[HTML]{00B050} 13.99} & \multicolumn{1}{c|}{{\color[HTML]{00B050} 94}}  & 50.00                 & 0                    & {\color[HTML]{C00000} 23.57} & {\color[HTML]{C00000} 29.66} & {\color[HTML]{C00000} 21.22} & {\color[HTML]{C00000} 9.93} \\
                                                                                              & \textbf{CodeLlama-Instruct-7B}      & {\color[HTML]{C00000} 13.38} & {\color[HTML]{C00000} 24.91} & \multicolumn{1}{c|}{65}                        & {\color[HTML]{C00000} 3.79}  & {\color[HTML]{C00000} 15.73} & \multicolumn{1}{c|}{{\color[HTML]{C00000} 29}} & {\color[HTML]{00B050} 4.84} & {\color[HTML]{C00000} 9.60}  & \multicolumn{1}{c|}{{\color[HTML]{C00000} 47}} & {\color[HTML]{C00000} 3.29} & {\color[HTML]{00B050} 14.81} & \multicolumn{1}{c|}{{\color[HTML]{00B050} 93}}  & 48.92                 & 20                    & {\color[HTML]{C00000} 28.77} & {\color[HTML]{00B050} 45.65} & {\color[HTML]{C00000} 21.13} & {\color[HTML]{C00000} 10.17} \\
                                                                                              & \textbf{CodeLlama-instruct-13B}     & {\color[HTML]{C00000} 13.28} & {\color[HTML]{C00000} 24.03} & \multicolumn{1}{c|}{{\color[HTML]{C00000} 62}} & {\color[HTML]{C00000} 6.14}  & {\color[HTML]{C00000} 15.16} & \multicolumn{1}{c|}{{\color[HTML]{C00000} 24}} & {\color[HTML]{C00000} 5.16} & {\color[HTML]{C00000} 11.20} & \multicolumn{1}{c|}{{\color[HTML]{C00000} 57}} & {\color[HTML]{C00000} 4.09} & {\color[HTML]{00B050} 15.72} & \multicolumn{1}{c|}{{\color[HTML]{00B050} 100}} & 44.38                 & 140                   & {\color[HTML]{00B050} 33.99} & {\color[HTML]{00B050} 50.60} & {\color[HTML]{C00000} 21.47} & {\color[HTML]{C00000} 10.08} \\
                                                                                              & \textbf{CodeLlama-Instruct-34B}     & {\color[HTML]{C00000} 1.89}  & {\color[HTML]{C00000} 3.77}  & \multicolumn{1}{c|}{{\color[HTML]{C00000} 11}} & {\color[HTML]{C00000} 1.11}  & {\color[HTML]{C00000} 4.77}  & \multicolumn{1}{c|}{{\color[HTML]{C00000} 10}} & {\color[HTML]{C00000} 4.29} & {\color[HTML]{C00000} 10.06} & \multicolumn{1}{c|}{{\color[HTML]{C00000} 54}} & {\color[HTML]{C00000} 4.74} & {\color[HTML]{00B050} 14.86} & \multicolumn{1}{c|}{{\color[HTML]{00B050} 88}}  & 49.68                 & 2                     & {\color[HTML]{00B050} 41.53} & {\color[HTML]{00B050} 50.79} & {\color[HTML]{C00000} 23.08} & {\color[HTML]{C00000} 10.80} \\
                                                                                              & \textbf{WizardCoder-Python-7B}      & {\color[HTML]{C00000} 8.00}  & {\color[HTML]{C00000} 20.12} & \multicolumn{1}{c|}{{\color[HTML]{C00000} 57}} & {\color[HTML]{C00000} 4.86}  & {\color[HTML]{C00000} 14.51} & \multicolumn{1}{c|}{{\color[HTML]{C00000} 24}} & {\color[HTML]{C00000} 3.25} & {\color[HTML]{C00000} 8.20}  & \multicolumn{1}{c|}{{\color[HTML]{C00000} 45}} & {\color[HTML]{00B050} 4.61} & {\color[HTML]{00B050} 15.60} & \multicolumn{1}{c|}{{\color[HTML]{00B050} 94}}  & 50.54                 & 0                     & /                            & {\color[HTML]{00B050} 55.50} & {\color[HTML]{C00000} 17.23} & {\color[HTML]{C00000} 13.63} \\
                                                                                              & \textbf{WizardCoder-Python-13B}     & {\color[HTML]{C00000} 12.44} & {\color[HTML]{C00000} 24.66} & \multicolumn{1}{c|}{{\color[HTML]{C00000} 65}} & {\color[HTML]{C00000} 5.21}  & {\color[HTML]{C00000} 18.25} & \multicolumn{1}{c|}{{\color[HTML]{C00000} 33}} & {\color[HTML]{00B050} 4.98} & {\color[HTML]{00B050} 11.22} & \multicolumn{1}{c|}{{\color[HTML]{00B050} 61}} & {\color[HTML]{00B050} 4.69} & {\color[HTML]{00B050} 16.51} & \multicolumn{1}{c|}{{\color[HTML]{00B050} 100}} & 47.62                 & 6                     & {\color[HTML]{00B050} 41.77} & {\color[HTML]{00B050} 62.19} & {\color[HTML]{C00000} 20.23} & {\color[HTML]{C00000} 14.91} \\
                                                                                              & \textbf{WizardCoder-Python-34B}     & {\color[HTML]{00B050} 15.88} & {\color[HTML]{00B050} 27.22} & \multicolumn{1}{c|}{72}                        & {\color[HTML]{C00000} 6.18}  & {\color[HTML]{C00000} 17.12} & \multicolumn{1}{c|}{{\color[HTML]{C00000} 27}} & {\color[HTML]{C00000} 4.79} & {\color[HTML]{C00000} 11.18} & \multicolumn{1}{c|}{{\color[HTML]{C00000} 58}} & {\color[HTML]{00B050} 6.54} & {\color[HTML]{00B050} 18.23} & \multicolumn{1}{c|}{{\color[HTML]{00B050} 105}} & 50.76                 & 0                     & {\color[HTML]{00B050} 44.94} & {\color[HTML]{00B050} 70.73} & {\color[HTML]{C00000} 22.02} & {\color[HTML]{C00000} 13.69} \\
\multirow{-7}{*}{\textbf{\begin{tabular}[c]{@{}c@{}}Instruction\\ Tuned\end{tabular}}}        & \textbf{WizardCoder-15B}            & {\color[HTML]{00B050} 14.41} & {\color[HTML]{C00000} 23.70} & \multicolumn{1}{c|}{{\color[HTML]{C00000} 64}} & {\color[HTML]{C00000} 5.00}  & {\color[HTML]{C00000} 18.84} & \multicolumn{1}{c|}{{\color[HTML]{C00000} 35}} & {\color[HTML]{C00000} 3.10} & {\color[HTML]{C00000} 11.62} & \multicolumn{1}{c|}{{\color[HTML]{C00000} 67}} & {\color[HTML]{C00000} 3.89} & {\color[HTML]{00B050} 15.94} & \multicolumn{1}{c|}{{\color[HTML]{00B050} 101}} & 33.15                 & 308                   & {\color[HTML]{00B050} 35.77} & {\color[HTML]{00B050} 58.12} & {\color[HTML]{C00000} 20.67} & {\color[HTML]{C00000} 8.36}  \\ \hline
                                                                                              & \textbf{Claude-1}                   & 21.55                        & 29.11                        & \multicolumn{1}{c|}{74}                        & 9.71                         & 17.77                        & \multicolumn{1}{c|}{28}                        & 1.20                        & 6.13                         & \multicolumn{1}{c|}{34}                        & 5.70                        & 16.56                        & \multicolumn{1}{c|}{95}                         & 47.95                     & 0                     & /                            & /                            & /                            & /                            \\
                                                                                              & \textbf{GPT-3.5-Turbo}              & 23.37                        & 39.67                        & \multicolumn{1}{c|}{102}                       & 12.18                        & 35.65                        & \multicolumn{1}{c|}{59}                        & 6.52                        & 13.39                        & \multicolumn{1}{c|}{71}                        & 9.31                        & 28.76                        & \multicolumn{1}{c|}{166}                        & 46.00                 & 78                    & /                            & 48.10                        & /                            & /                            \\
\multirow{-3}{*}{\textbf{\begin{tabular}[c]{@{}c@{}}Close\\ Source\end{tabular}}}             & \textbf{GPT-4}                      & \textbf{30.52}               & \textbf{42.94}               & \multicolumn{1}{c|}{\textbf{110}}              & \textbf{24.18}               & \textbf{45.72}               & \multicolumn{1}{c|}{\textbf{72}}               & 6.66                        & \textbf{15.76}               & \multicolumn{1}{c|}{\textbf{83}}               & \textbf{18.76}              & \textbf{38.29}               & \multicolumn{1}{c|}{\textbf{203}}               & \textbf{52.16}        & 0                     & /                            & 67.00                        & /                            & /                            \\ \hline
\end{tabular}
\end{adjustbox}
\label{tab:eval-coderucb}
\end{table*}

To explore whether including program context is beneficial for code LLMs in solving programming tasks, we conducted experiments comparing basic prompts augmented with program context against standard few-shot prompts~\cite{fewshot}. These experiments were applied across four different tasks within the CoderUJB. Specifically, the program context prompts for the functional code generation task, the code-based test generation task, and the automatic program repair task are the same as those presented in Section~\ref{sec:3.4}. For defect detection tasks, the program context prompts were designed based on the prompt of functional code generation. The few-shot prompt examples were chosen from filter-out samples when creating CoderUJB or from other benchmarks~\cite{quixbugs} of the same task. It is important to note that we did not include an issue-based test generation task as prior research~\cite{issuetest} suggests that specific program contexts are not applicable in such scenarios. Due to the page limit, we conducted experiments on two representative open-source code LLMs, CodeLlama-13B and StarCoder-Base-15B, while placing experiments evaluating more LLMs in subsequent sections.


Table~\ref{tab:eval-prompt} shows the results of the prompt comparison experiments. We can find that program context is helpful for most programming tasks. Specifically, in functional code generation and code-based test generation, the program context prompt outperformed the few-shot prompt. For example, StarCoderBase-15B scored 15.32 and 75 in $pass\text{-}all@k\text{=}1$ and $count\text{-}all@n\text{=}20$, outperforming the results from other few-shot prompts. Moreover, in code-based test generation, tests generated with program context demonstrated significantly better coverage than those from few-shot prompts. In the automated program repair task, we observed notable improvements in StarCoderBase-15B when using the program context prompt. Therefore, such comparison results clearly show the good quality of the basic prompts from CoderUJB and highlight the value of supplying full program contexts (i.e., all source code and execution environment) in CoderUJB. 

However, in the defect detection task, we found that the $accuracy$ of prompt with program context is slightly lower than the few-shot prompt. For example, the $accuracy$ of CodeLlama-13B with the program context prompt was 48.60\%, lower than the 49.68\% and 51.51\% achieved with the few-shot prompt. We then further show the $error\text{-}count$ metric of the number of answers that could not be parsed correctly (i.e., not a valid answer), which reveals that the code LLMs may fail to generate valid answers when using program context prompt, ultimately leading to lower $accuracy$. To capture the impact of invalid answer, we will report both $accuracy$ and $error\text{-}count$ for subsequent experiments. Nevertheless, we are still choosing the program context prompt as the basic prompt for the defect detection task because it achieves similar accuracy to the few-shot prompt, as it also contains two few-shot examples. Additionally, we believe it may perform even better in subsequent experiments since it contains additional program context. 

Note that CoderUJB provides a base prompt design that is as reliable as possible. We recognize that these base prompt designs are not best practices. However, this is exactly why we introduced CoderUJB as a comprehensive context (i.e., all source code and execution environment) benchmark. We encourage other researchers to use this benchmark to explore and create even better prompt designs, as there is already a lot of interesting work~\cite{codeagent1,codeagent2,codeagent3} in this field, and CoderUJB can offer a comprehensive and fair framework for such research.

\mybox{Conclusion 1: The program context is useful for functional code generation, code-based test generation, and automated program repair.}


\subsubsection{RQ2: How Do Open-Source and Closed-Source LLMs Perform Under CoderUJB}

\label{sec:rq1}
\begin{figure}
  \centering

  \includegraphics[width=0.7\columnwidth]{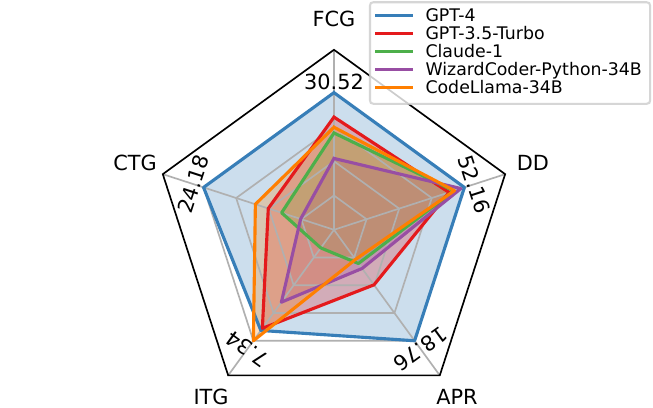}
  \caption{Evaluation results ($accuracy$ for DD and $pass\text{-}all@k\text{=}1$ for others) of open-source LLMs and closed-source LLMs under CoderUJB.\remove{The metric for the Defect Detection (DD) task is $accuracy$, while the evaluation metric for all other tasks is $pass\text{-}all@k\text{=}1$.}}
  \label{fig:eval_rq1}
\end{figure}

Table~\ref{tab:eval-coderucb} shows the evaluation results of the selected LLMs in five programming scenarios on key performance metrics (i.e., $pass\text{-}all@k$, $count\text{-}all@n$, $accuracy$).
To better demonstrate the performance differences between open-source and closed-source LLMs, we have exhibited the comparison using $pass\text{-}all@k$ and $accuracy$ radar plots in Figure~\ref{fig:eval_rq1}, featuring three leading open-source code LLMs and three closed-source LLMs under CoderUJB. 

After combining detailed experimental results from Table~\ref{tab:eval-coderucb} with Figure~\ref{fig:eval_rq1}, it can be found that current LLMs fail to achieve the same impressive results as HumanEval~\cite{codellama} and CoderEval~\cite{codereval} under CoderUJB. Specifically, the most powerful open-source coder LLMs CodeLlama-34B and closed-source LLM GPT-4 can only achieve $pass\text{-}all@k\text{=}1$ metrics of 22.82 and 30.52 under the functional code generation task, much lower than their results of 45.11~\cite{codellama} and 67.00~\cite{gpt4report} on HumanEval. Their performance under other programming tasks is even worse than that under the functional function generation task, e.g., pass-all@k=1 of GPT-4 under the other 3 code generation tasks are only 24.18, 15.76, and 18.76. Moreover, the CodeShell-7B model, which utilizes limited training resources, falls short in delivering satisfactory results compared to its performance in Humaneval. This discrepancy stems from its training corpus, which is biased towards simplistic code samples and relies on significantly fewer resources. Such results underscore the value of challenging benchmarks such as CoderUJB. Therefore, we conclude that CoderUJB provides much more challenging programming questions than HumanEval and CoderEval. Also, the other programming tasks are more complicated than the functional code generation task that previous studies~\cite{codex,codegen,codereval,mbpp} mainly focused on because the requirements of the other tasks are more abstract, requiring a deeper understanding and the ability to address more complex and varied programming situations~\cite{chatteser,issuetest,defects4j,aprplm}.

In addition, It is also important to note that none of the current LLMs could achieve acceptable results in defect detection tasks, highlighting the formidable challenge of this task. Specifically, even GPT-4 achieves only a 52.16\% accuracy rate, which is marginally better than random guessing. Previous studies~\cite{chatgptcewdetect1,chatgptcewdetect2} have also found that models like ChatGPT perform inadequately when detecting common weaknesses enumerated (CWE) vulnerabilities. We believe one critical issue is that most of the code defects within CoderUJB are complex logic errors (i.e., errors producing unintended behaviors)~\cite{logicalerror} rather than syntax or API misused errors. To pinpoint such logic errors in code, a model would need an extensive grasp of the entire project, which is also challenging for experienced developers. Further complicating matters is that defect detection is a classification task~\cite{codexglue}. This substantially differs from the mainstream pre-training tasks, i.e., autoregressive generation~\cite{gpt1,bart}. Thus, the currently employed decoder-only autoregressive LLM suffers from inherent disadvantages when dealing with classification tasks, and such a conclusion can also be found in other natural language classification tasks~\cite{bert,llmclassification}. Given the significant challenge that defect detection poses to current LLMs, this study will not analyze the results of this task in depth. Instead, we calls for researchers to concentrate their efforts on improving the defect detection capabilities of LLMs.


\mybox{Conclusion 2: In basic question-and-answer or completing scenarios, current LLMs have not achieved satisfactory results in CoderUJB representing real programming challenges, especially in the defect detection task, where all LLMs are almost randomly guessing.}

Next, we compare the results of open-source LLMs and closed-source commercial LLMs to quantify the gap between the two types of models. It can be observed that the performance comparison results of the two types of LLMs differ under different programming tasks. In the area of functional code generation, the top-tier open-source LLMs (i.e., CodeLlama-34B) manage to match the performance of the well-performing closed-source model (i.e., GPT-3.5-Turbo~\cite{chatgpt}).
On the two test generation tasks, top open-source LLMs (i.e., CodeLlama-34B) even surpass the GPT-3.5-Turbo. We believe that GPT-3.5-Turbo's performance may have been affected by instruction fine-tuning, which we will investigate further in Section~\ref{sec:rq4}.
When it comes to automated program repair tasks, there's still a noticeable performance gap between the best open-source LLMs and excellent closed-source generic LLMs. The pass-all@k metrics for GPT-3.5-Turbo are 9.31 and 28.76, outperforming the corresponding metrics (6.54 and 18.32) for the top open-source LLM, WizardCoder-Python-34B.


\mybox{Conclusion 3: Advance open-source LLMs have made significant progress, achieving similar or even better performance than the excellent closed-source model GPT-3.5-Turbo on the functional code generation task and the two test generation tasks, but perform poorly on the automated program repair. Meanwhile, GPT-4 surpasses all other LLMs substantially, suggesting that scaling remains a powerful tool for enhancing model performance.}


\subsubsection{RQ3: How Does Continued Pre-Training of Specific Programming Language (PL) Data Affect the Performance of Code LLMs Under CoderUJB}
\label{sec:rq3}
To address this research question, we look at how the Base LLMs listed in Table~\ref{tab:eval-coderucb} perform compared to Specific-PL-Base LLMs under CoderUJB. In order to show the comparison results of the two classes of LLMs more intuitively, we have highlighted the results where Specific-PL-Base LLMs fall short of their Base counterparts in {\color[HTML]{C00000} red} and vice versa in {\color[HTML]{00B050} green}. We can find that the effects of specific programming language (PL) continued pre-training can vary greatly depending on the specific task. 

For CoderUJB-FCG, HumanEval, and CoderEval, all of which are functional code generation tasks. Specific PL training can boost the performance of corresponding PL tasks and hinder other PL tasks. For example, CodeLlama-Python and StarCoder-Python generally perform better in HumanEval-Py and CoderEval-Py but worse in CoderEval-Java and CoderUJB-FCG (Java tasks). On the other hand, StarCoder-Java performs better in Java tasks but worse in Python tasks. This phenomenon is consistent with the consensus of researchers, i.e., in-domain training enhances the performance of in-domain tasks while potentially hindering the performance of tasks outside the domain~\cite{fine-tune-bert1,fine-tune-bert2,fine-tune-bert3}.
However, things get different when it comes to more challenging tasks, i.e., CoderUJB-ITG and APR. The performance influence due to Specific PL training is random and less substantial in the case of those tasks when compared with functional code generation tasks. For example, we can observe two counter-instances from CoderUJB-ITG (Java tasks), i.e., StarCoder-Python getting better results in ITG after Python training while StarCoder-Java getting worse in ITG after Java training. And we can find more counter-instances in CoderUJB-APR. 

We attribute this to the fact that test generation and automatic program repair tasks are more different from the pre-training task compared with functional code generation. In other words, When the downstream task is more similar to the pre-training task, such as in the case of functional code generation, the performance boost or decline is more predictable and substantial. On the other hand, when the downstream task is substantially different from the pre-training task, such as in the case of automated program repair, the effect of specific PL training tends to be unpredictable.

Furthermore, the varied outcomes across different tasks emphasize CoderUJB's value as a comprehensive evaluation benchmark that incorporates a range of programming challenges.


\mybox{Conclusion 4: The impact of specific PL training might relate to how much the downstream task differs from the pre-training task. The more similar the task (e.g., functional code generation), the more predictable and substantial the performance impact (i.e., boost the performance of corresponding PL tasks and hinder other PL tasks). Conversely, if the task differs more substantially (e.g., automated program repair), the effect due to specific PL training tends to be unpredictable.}

\subsubsection{RQ4: How Does Instruction Fine-Tuning Influence the Performance of Code LLMs in CoderUJB}
\label{sec:rq4}

Finally, we assessed the performance of Instruction-Tuned LLMs compared to their original counterparts, as shown in Table~\ref{tab:eval-coderucb}. For a more intuitive comparison, we have highlighted the results where Instruction-Tuned LLMs fall short of their counterparts base models in {\color[HTML]{C00000} red} and vice versa in {\color[HTML]{00B050} green}. The results indicated that instruction tuning can yield vastly different performance impacts across diverse programming tasks. 

Specifically, most Instruction-Tuned LLMs struggled to outperform their base models when it came to functional code generation and test generation tasks. For instance, CodeLlama-Instruct-13B scored lower in the $pass\text{-}all@k$ metrics during the function code generation task, with a 39.39\% ($k\text{=}1$) and 31.05\% ($k\text{=}10$) drop respectively compared to its base model (CodeLlama-13B). This performance drop was consistent among most open-source Instruction-Tuned LLMs. Such a result was different with HumanEval where instruction fine-tuning largely increased its performance. We believe that the simplicity of HumanEval's single-function generation tasks does not reflect the complexity of real-world development scenarios, a gap that CoderUJB addresses with its features. This distinction is why instruction fine-tuning has varying impacts between the two, and underscores the value of CoderUJB as a practical programming evaluation benchmark. 


\begin{table}[]
\caption{$pass\text{-}syntax@k\text{=}1$ and $pass\text{-}compile@k\text{=}1$ (denoted as $p\text{-}s@1$ and $p\text{-}c@1$) results for CoderUJB.}
\begin{adjustbox}{width=1.0\columnwidth}
\begin{tabular}{|c|c|cc|cc|cc|}
\hline
\multirow{2}{*}{\textbf{Model-Type}}                                                  & \multirow{2}{*}{\textbf{Model-ID}} & \multicolumn{2}{c|}{\textbf{FCG}}                                      & \multicolumn{2}{c|}{\textbf{CTG}}                                      & \multicolumn{2}{c|}{\textbf{ITG}}                                      \\ \cline{3-8} 
                                                                                      &                                    & \multicolumn{1}{c}{\textbf{$p\text{-}s@1$}} & \textbf{$p\text{-}c@1$} & \multicolumn{1}{c}{\textbf{$p\text{-}s@1$}} & \textbf{$p\text{-}c@1$} & \multicolumn{1}{c}{\textbf{$p\text{-}s@1$}} & \textbf{$p\text{-}c@1$} \\ \hline\hline
\multirow{3}{*}{\textbf{Base}}                                                        & \textbf{CodeLlama-7B}              & \multicolumn{1}{c}{92.63}                   & 69.58                   & \multicolumn{1}{c}{70.64}                   & 36.71                   & \multicolumn{1}{c}{70.14}                   & 42.36                   \\ 
                                                                                      & \textbf{CodeLlama-13B}             & \multicolumn{1}{c}{94.14}                   & 63.49                   & \multicolumn{1}{c}{70.54}                   & 37.89                   & \multicolumn{1}{c}{82.29}                   & 55.43                   \\ 
                                                                                      & \textbf{CodeLlama-34B}             & \multicolumn{1}{c}{96.16}                   & 61.58                   & \multicolumn{1}{c}{77.11}                   & 39.32                   & \multicolumn{1}{c}{87.86}                   & 51.87                   \\ \hline
\multirow{3}{*}{\textbf{\begin{tabular}[c]{@{}c@{}}Instruction\\ Tuned\end{tabular}}} & \textbf{CodeLlama-Instruct-7B}     & \multicolumn{1}{c}{90.53}                   & 40.65                   & \multicolumn{1}{c}{88.64}                   & 18.04                   & \multicolumn{1}{c}{89.39}                   & 33.34                   \\ 
                                                                                      & \textbf{CodeLlama-instruct-13B}    & \multicolumn{1}{c}{95.99}                   & 40.76                   & \multicolumn{1}{c}{80.61}                   & 21.57                   & \multicolumn{1}{c}{89.29}                   & 32.06                   \\ 
                                                                                      & \textbf{CodeLlama-Instruct-34B}    & \multicolumn{1}{c}{11.30}                   & 6.43                    & \multicolumn{1}{c}{11.54}                   & 3.89                    & \multicolumn{1}{c}{90.51}                   & 34.28                   \\ \hline
\end{tabular}
\end{adjustbox}
\label{tab:eval-syntax}
\end{table}

To investigate the cause of performance degradation, we further show the $pass\text{-}syntax@k\text{=}1$ and $pass\text{-}compile@k\text{=}1$ metrics for the three code generation tasks in Table~\ref{tab:eval-syntax}. Interestingly, we found that the syntactical correctness in the code generated by the Instruction-Tuned LLMs, e.g., CodeLlama-Instruct-7B and 13B, was similar to their corresponding base models. Therefore, the decline in their performance is likely due to the lower quality of the code solutions they generate (e.g., lower $pass\text{-}compile@k$ and $pass\text{-}all@k$ scores) and not because of rejected answers or answers that cannot be parsed. Meanwhile, CodeLlama-Instruct-34B exhibited a noteworthy drop in $pass\text{-}syntax@k\text{=}1$ and $pass\text{-}all@k$ during FCG and CTG. This decline is attributed to mode collapse~\cite{modecollapse}, characterized by the generation of identical solutions for all prompts, and it occurs only in one model, leading us to disregard these results as invalid in our analysis. Thus, we believe that a possible reason is the high similarities between pre-training tasks, functional code, and test generation tasks, so that base LLMs can accomplish these tasks without aligning with upstream and downstream tasks~\cite{alignment1,alignment2,alignment3}. Therefore, they are able to leverage the full capabilities of the base models. On the contrary, instruction tuning, with its varied form, might result in disturbances when using such diverse LLMs for direct code generation tasks. 


Conversely, for the automated program repair task, instruction fine-tuning actually enhanced model performance. This trend applied to most instruction fine-tuned LLMs. We believe this is due to the significantly different format of the automated program repair task compared to the pre-training task. The differences in upstream and downstream tasks lead to poor performance when directly applying Base LLMs. However, instruction tuning enhances the model's adaptability and applicability to diverse tasks through data in diverse task formats~\cite{flan}, ultimately improving the performance of LLMs for automated program repair tasks. 



\mybox{Conclusion 5: Instruction tuning reduces the performance of LLMs under tasks highly consistent with the pre-trained task (e.g., functional code generation and test generation tasks), while boosting the performance of tasks that starkly differ from the pre-training tasks' format (e.g., automated program repair). Lastly, we encourage further exploration and studies to uncover more effective fine-tuning strategies for LLMs.}

\section{Implications and Discussions}
Our study reveals the following important practical guidelines for future research on LLMs of software engineering.

\parabf{Program context is important.}
The findings from RQ1 imply that incorporating basic program context can enhance performance across various programming tasks. Consequently, we encourage researchers to investigate and devise advanced prompt designs and methods to fully harness the potential of program context.

\parabf{Caution with specific programming language continued pre-training.}
The results of RQ3 reveal that for tasks similar to those used in pre-training, specific PL focused training generally enhances performance in the corresponding language while potentially impeding performance in others. Conversely, if the task differs more substantially from the pre-training task, the effect due to specific PL training tends to be unpredictable. Therefore, researchers should carefully balance the training between language-specific tasks and those in other PL to determine the extent and volume of data appropriate for further pre-training.

\parabf{Caution with instruction fine-tuning.}
The results of RQ4 indicate that instruction fine-tuning reduces the performance of LLMs under tasks highly consistent with the pre-trained task. For such tasks, we suggest researchers use the original base LLMs as the foundation for their application. Conversely, for tasks that starkly differ from the pre-training tasks, instruction-tuned LLMs tend to perform better and should be considered. Lastly, we encourage further studies to uncover more effective fine-tuning strategies.

\parabf{More extensive evaluations are needed.}
The conclusions drawn from RQ3 and RQ4 suggest that varying programming tasks can lead to disparate results when employing the same training strategy. Consequently, we advocate that researchers should assess their LLMs and techniques using a more comprehensive benchmark such as CoderUJB to obtain more reliable evaluation outcomes.

\section{Threats to Validity}
\parabf{Threats to Internal Validity.}
The threats to internal validity mainly lie in the potential bugs in our implementation. To mitigate these risks, the authors have meticulously reviewed the code and scripts. Furthermore, we have released the code, scripts, and all generated results for public scrutiny at~\cite{coderujbgithub}.

\parabf{Threats to External Validity.}
This threats mainly lie in the LLMs adopted in this study. To address these concerns, we have conducted an extensive literature review and believe that the selected LLMs are sufficiently representative and influential within this field. 

\parabf{Threats to Construct Validity.}
The threats to construct validity in our study primarily arise from the metrics used in our evaluations. To mitigate these threats, we initially adopted the widely accepted $pass@k$ metric and verified that each coding problem was accompanied by adequate test coverage. Additionally, we utilized a range of widely-recognized metrics, specifically $accuracy$, $count@n$, and $coverage@n$, to provide a comprehensive evaluation of the models.

\section{CONCLUSIONS}
CoderUJB is introduced as a benchmark that advances the evaluation of large language models (LLMs) by simulating real-world software engineering tasks with executable code extracted from 17 open-source Java projects. Our study delved into the performance of LLMs, highlighting difficulties in non-functional code generation and defect detection tasks.
The research revealed the delicate balance required when continuing pre-training and instruction fine-tuning, as they can inadvertently decrease performance in certain scenarios. These observations suggest that a nuanced approach to training LLMs is essential to ensure versatility and robustness across various coding tasks.
In essence, CoderUJB contributes to setting more exacting benchmarks for assessing LLMs in software engineering and provides insights into the complexities of model training, guiding future research toward developing more refined and adaptable LLMs for practical coding applications.

\balance
\bibliographystyle{ACM-Reference-Format}
\bibliography{ref}

\end{document}